\documentclass[aps,prl,reprint,groupedaddress]{revtex4-2}

\usepackage{amsmath}
\usepackage{amssymb}
\usepackage{graphicx}

\begin{document}


\title{The Quantum Transition State}%

\author{Pouya Khazaei}%
 \email{pkhazaei@umich.edu}
\affiliation{Department of Chemistry, University of Michigan, Ann Arbor, MI 48109, U.S.A.}
\affiliation{Chemical Physics Theory Group, Department of Chemistry, University of Toronto, Toronto, Ontario M5S 3H6, Canada}
\affiliation{Department of Physical and Environmental Sciences, University of Toronto Scarborough, Toronto, Ontario M1C 1A4, Canada}

\date{\today}

\begin{abstract}
For nearly a century, the transition state---the structure organizing motion
between reactants and products---has seemed to lack an exact quantum
counterpart. Its geometry---a recrossing-free dividing surface whose one-way
flux gives the exact rate---seems to require position and momentum direction at
once, in tension with the uncertainty principle. The obstruction seems
fundamental. It is not. The exact quantum equations generate a phase-space flow
whose stable and unstable manifolds intersect in a unique bounded
trajectory---the quantum transition state---that anchors such a surface.
Quantum mechanics does not destroy the transition state; it becomes quantum.
\end{abstract}
\maketitle
Since its formulation in the 1930s by Eyring, Polanyi, Wigner, and others
\cite{Eyring_Polanyi_1931,Pelzer_Wigner_1932,Eyring_1935,Eyring_1938,Eyring_1938_1,Wigner_1938,Evans_Polanyi_1935,miller_spiers_1998},
transition-state theory (TST) has organized reactivity around the transition
state, a sharply defined bottleneck separating reactants from products. This
statistical picture was later given an exact phase-space realization
\cite{Waalkens_Schubert_Wiggins_2008}. In autonomous systems, a periodic orbit
anchors the dividing surface in two degrees of freedom
\cite{Pechukas_McLafferty_1973,pollak_transition_1978}, while a normally
hyperbolic invariant manifold (NHIM) does so in higher dimensions
\cite{wiggins_normally_1994,uzer_geometry_2002,wiggins_role_2016}. Their
nonautonomous counterparts are a hyperbolic transition-state trajectory in one
dimension and a time-dependent normally hyperbolic center manifold in higher
dimensions \cite{bartsch_transition_2005,rice_time-dependent_2008}. These
invariant structures comprise trajectories trapped near the barrier and anchor
locally recrossing-free dividing surfaces; their stable and unstable manifolds
separate reactive from nonreactive motion. In the absence of global returns,
each reactive trajectory crosses once, and the resulting one-way flux gives the
exact rate without a recrossing correction.

Quantum mechanics appeared to erase this geometry. Wigner identified the obstruction: defining a one-way flux appears to require
specifying both position on the dividing surface and momentum direction, in
tension with the uncertainty principle
\cite{wigner_crossing_1932,hirschfelder_quantum-mechanical_1939}. This
obstruction split subsequent quantum rate theory between two properties united
by the classical transition state: an exact rate and a dynamically privileged
dividing surface
\cite{Truhlar_Garrett_Klippenstein_1996,pollak_reaction_2005,Jang_Voth_2016,Jang_Voth_2017}.
Flux-correlation approaches retained the exact rate by incorporating recrossing
dynamically, rendering the dividing surface immaterial
\cite{yamamoto_quantum_1960,miller_quantum_1983,miller_beyond_1993,Topaler_Makri_1993,Topaler_Makri_1994,Geva_Shi_Voth_2001,predescu_optimal_2005}.
Quantum TST approaches instead retained a privileged surface, selected through
an effective potential, free energy, or action, but yielded an approximate rate
\cite{miller_quantum_1974,McLafferty_Pechukas_1974,Voth_Chandler_Miller_1989,pollak_new_1998,Jang_Voth_2000,Shi_Geva_2002,Craig_Manolopoulos_2005,
Schubert_Waalkens_Wiggins_2006,Richardson_Althorpe_2009,Richardson_2016,hele_derivation_2013}.
Closest to the present work, Hyland and Martens computed a quantum-trajectory
decay rate through the reactive segment of the classical energy separatrix,
recovering the correct tunneling scaling perturbatively
\cite{hyland_toward_2014}. What remained missing was the geometry itself: a
recrossing-free dividing surface generated by the exact quantum equations of
motion, whose instantaneous flux directly yields the exact rate
\cite{Truhlar_Garrett_Klippenstein_1996,pollak_quantum_1993,small_quantifying_2006}.

Here we show that such an object exists. We construct it for a one-dimensional
barrier top. The apparent obstruction rests on a single assumption: that
one-way flux requires simultaneous position and momentum information. It does
not. The exact quantum equations generate a phase-space flow whose stable and
unstable manifolds intersect in a unique bounded trajectory---the quantum
transition state. This trajectory anchors a recrossing-free dividing surface
across which probability flows strictly one way. Quantum mechanics does not
abolish the transition state; it changes its form.

Background on the dynamical-systems and quantum-trajectory framework is
provided in the Supplemental Material (SM)~\cite{supplemental_material}.

To construct these structures, we use the trajectory formulation of quantum
mechanics
\cite{bohm_suggested_1952,bohm_suggested_1952-1,Durr_Teufel_2009,Holland_1993,Norsen_2017,Bricmont_2016}.
Trajectory-based formulations of quantum dynamics have been widely
applied in chemical dynamics
\cite{Wyatt_Trahan_2005,garashchuk_semiclassical_2002,
Garashchuk_Rassolov_2003,Donoso_Martens_1998,donoso_quantum_2001,
maddox_quantum_2001,maddox_quantum_2002,Sanz_Borondo_2007,
sanz_contextuality_2009,Goldfarb_Degani_Tannor_2006,liu_monte_2004,
liu_bohms_2005}
and analyzed from a dynamical-systems perspective
\cite{Wisniacki_Pujals_2005,Wisniacki_Pujals_Borondo_2006,
Borondo_Luque_Villanueva_Wisniacki_2009}.
Let $\mathcal Q=\mathbb R$ be the configuration space and
$\mathcal P=T^*\mathcal Q\simeq\mathbb R^2$ its phase space, with
phase-space points $X=(x,p)^T$. Writing $\psi=Re^{iS/\hbar}$ \cite{madelung_quantentheorie_1927} recasts the
Schr\"odinger equation, without approximation, as a continuity equation
coupled to a quantum Hamilton--Jacobi equation, with phase-space dynamics
\begin{equation}
\dot x=\frac{p}{m},
\qquad
\dot p=-\nabla(V+Q),
\label{eq:dbb}
\end{equation}
where $Q=-(\hbar^2/2m)\nabla^2R/R$ is the quantum potential. We denote the
resulting flow on $\mathcal P$ by $\Theta_{t,t_0}$.

Eq.~\eqref{eq:dbb} defines a trajectory through every point of
$\mathcal P$, but
$\psi$ endows only a subset of them with
probability. Its phase fixes the momentum, $p=\nabla S(x,t)$, singling out
the graph
\begin{equation}
  \Gamma_\psi(t)=\bigl\{(x,p)\in\mathcal P : p=\nabla S(x,t)\bigr\},    
\end{equation}
which is invariant under the same flow,
$\Theta_{t,t_0}\bigl(\Gamma_\psi(t_0)\bigr)=\Gamma_\psi(t)$
\cite{deLeon_Sardon_2017,RomanRoy_2021}: a trajectory launched on the graph
never leaves it. Restricted to $\Gamma_\psi(t)$, Eq.~\eqref{eq:dbb} collapses
to the first-order law $\dot x=v(x,t)$, whose solutions are the
characteristics of the continuity equation
\begin{equation}
  \partial_t\rho+\nabla\cdot(\rho v)=0,
  \qquad
  \rho=|\psi|^2,
  \qquad
  v=\frac{\nabla S}{m},
  \label{eq:cont}
\end{equation}
so the flow transports the Born probability exactly
\cite{Durr_Goldstein_Zanghi_1992}. The associated phase-space measure,
\begin{equation}
  d\mu_t=\rho(x,t)\,\delta\!\bigl(p-\nabla S(x,t)\bigr)\,dx\,dp,
\end{equation}
is concentrated by the delta function on that graph,
$\operatorname{supp}\mu_t\subseteq\Gamma_\psi(t)\subset\mathcal P$
\cite{wiggins_geometric_2026}. A recrossing is therefore an event on
$\Gamma_\psi(t)$: a characteristic meets the
moving dividing surface $\Sigma(t)$ more than once.

As in classical TST
\cite{Martens_2002,rice_time-dependent_2008,bartsch_transition_2005},
we construct the transition-state geometry near the maximum of
$U_{\mathrm{eff}}=V+Q$. With the bare barrier top at the origin, let
$x_*(t)=\arg\max_xU_{\mathrm{eff}}(x,t)$. Expanding about $x_*(t)$ to second order gives
\begin{align}
\dot X&=A(t)X+F(t), \qquad X=(x,p)^T,\notag\\
A(t)&=
\begin{pmatrix}
0&1/m\\
m\Omega^2(t)&0
\end{pmatrix},
\qquad
F(t)=
\begin{pmatrix}
0\\ f(t)
\end{pmatrix}.
\label{eq:inhomo}
\end{align}
Here $m\Omega^2(t)=-\partial_x^2U_{\mathrm{eff}}(x_*(t),t)>0$ for a
nondegenerate maximum, and $f(t)=-m\Omega^2(t)x_*(t)$ measures its
displacement from the origin. For a Gaussian wavepacket on a harmonic
barrier, the expansion is exact, with
$\Omega^2(t)=\omega^2+a(t)/m$ and
$x_*(t)=a(t)q_t/[m\Omega^2(t)]$.

In autonomous one-degree-of-freedom TST, the transition state is a hyperbolic
saddle equilibrium fixed at the barrier top. Its linearization has real
eigenvalues of opposite sign, with corresponding eigendirections tangent to the
stable and unstable manifolds. For a time-dependent barrier, the equilibrium is
replaced by a bounded trajectory whose stable and unstable directions are
invariant subspaces of the propagator, rather than instantaneous eigenspaces of
$A(t)$. This global splitting is an \emph{exponential dichotomy} \cite{coppel_dichotomies_1978,barreira_stability_2008}. If
$\Phi(t,\tau)$ is the propagator of $\dot X=A(t)X$, a dichotomy on the time interval
$\mathbb R$ consists of projections $P(t)$ and constants $K,\alpha>0$
satisfying
\begin{equation}
\Phi(t,\tau)P(\tau)=P(t)\Phi(t,\tau)
\end{equation}
and
\begin{align}
\|\Phi(t,\tau)P(\tau)\|
&\le Ke^{-\alpha(t-\tau)}, && t\ge\tau,\\
\|\Phi(t,\tau)[I-P(\tau)]\|
&\le Ke^{-\alpha(\tau-t)}, && t\le\tau.
\end{align}
Thus
$\mathcal S^s(t)=\operatorname{Ran}P(t)$ decays forward and
$\mathcal S^u(t)=\ker P(t)$ decays backward. In the present two-dimensional
system, both are one-dimensional and
$\mathbb R^2=\mathcal S^s(t)\oplus\mathcal S^u(t)$.

\emph{Central result.---}
If the homogeneous system admits an exponential dichotomy and $F(t)$ is
bounded, Eq.~\eqref{eq:inhomo} has a unique trajectory
$X_b(t)=(x_b(t),p_b(t))^T$ bounded for all $t\in\mathbb R$
\cite{coppel_dichotomies_1978}. This \emph{quantum transition-state
trajectory} anchors the affine stable and unstable slices
\begin{align}
W_t^s(X_b)&=X_b(t)+\mathcal S^s(t),\notag\\
W_t^u(X_b)&=X_b(t)+\mathcal S^u(t),
\label{eq:affine-manifolds}
\end{align}
whose points approach $X_b$ in forward and backward time, respectively.
Their unique intersection is $X_b(t)$.

The projection $x_b(t)$ anchors the moving dividing surface
\begin{equation}
\Sigma(t)=\{(x,p)\in\mathcal P:x=x_b(t)\}.
\label{eq:surface}
\end{equation}
When $W_t^s(X_b)$ and $W_t^u(X_b)$ meet the physical graph
$\Gamma_\psi(t)$ on opposite sides of $\Sigma(t)$, their intersections bound
a middle interval whose characteristics cross $\Sigma(t)$ exactly once; the
exterior characteristics never cross. All crossings have the same orientation,
so $\Sigma(t)$ is recrossing-free and carries one-way probability current.

We establish the dichotomy on $\mathbb R$ in two steps: first on the
half-lines $\mathbb R_\pm$, then globally by excluding nonzero bounded
homogeneous solutions \cite{ju_roughness_2001}. Exponential dichotomies
persist under sufficiently small perturbations
\cite{coppel_dichotomies_1978}. In particular, Corollary~3.1 of
Ref.~\citenum{ju_roughness_2001} and its $\mathbb R_-$ analogue imply that if
$A(t)=B(t)+C(t)$, $B(t)$ admits an exponential dichotomy, and
$\limsup_{|t|\to\infty}\|C(t)\|$ is sufficiently small, then $A(t)$ inherits
half-line dichotomies with the same stable and unstable dimensions.

For the Gaussian barrier-top model,
$A(t)=A_\infty+\Delta A(t)$, where
\begin{equation}
A_\infty=
\begin{pmatrix}
0 & 1/m\\
m\omega^2 & 0
\end{pmatrix}.
\label{eq:Ainfty}
\end{equation}
The eigenvalues $\pm\omega$ make $\dot X=A_\infty X$ hyperbolic and hence
exponentially dichotomic. Since
$\Omega^2(t)=\omega^2+a(t)/m$ and $a(t)\to0$ as $|t|\to\infty$,
$\|\Delta A(t)\|\to0$. The system $\dot X=A(t)X$ therefore inherits
dichotomies on both half-lines, with one-dimensional stable and unstable
subspaces.

It remains to exclude nonzero bounded homogeneous solutions. If
$X_h=(x_h,p_h)^T$ were such a solution, then the bounded function
$y=x_h^2$ would satisfy
\begin{equation}
\ddot y=2\dot x_h^{\,2}+2\Omega^2(t)x_h^2\ge0.
\end{equation}
Thus $y$ would be convex and bounded above on $\mathbb R$, and hence constant.
Since $\Omega^2(t)>0$, this implies $x_h\equiv0$ and
$p_h=m\dot x_h\equiv0$. Therefore the forward stable subspace
$\mathcal S_+^s(t)$ and backward unstable subspace $\mathcal S_-^u(t)$ have
trivial intersection. Being one-dimensional subspaces of $\mathbb R^2$, they
are complementary, so the half-line dichotomies combine into an exponential
dichotomy on $\mathbb R$ \cite{ju_roughness_2001,dieci_exponential_2010}.

With bounded forcing $F(t)$, the unique bounded solution is then
\cite{coppel_dichotomies_1978}
\begin{align}
X_b(t)
&=\int_{-\infty}^t\Phi(t,\tau)P(\tau)F(\tau)\,d\tau \notag\\
&\quad-\int_t^\infty\Phi(t,\tau)[I-P(\tau)]F(\tau)\,d\tau .
\label{eq:bounded}
\end{align}

Relative to the dividing surface anchored by $x_b(t)$, the reactant population
is
\begin{equation}
N(t)=\int_{x<x_b(t)}\rho(x,t)\,dx,
\qquad
\rho=|\psi|^2.
\end{equation}
The continuity equation gives the probability flux relative to the moving
surface,
\begin{equation}
-\dot N(t)=j(x_b(t),t)-\rho(x_b(t),t)\dot x_b(t)
\equiv J_\Sigma(t),
\label{eq:flux}
\end{equation}
where the second term accounts for surface motion. Because $\Gamma_\psi(t)$ assigns a unique momentum $p=\partial_xS$ to each
$x$, the physical flow is monokinetic. Using
$j=\rho\,\partial_xS/m$ \cite{Daumer_Durr_Goldstein_Zanghi_1997,Cohen_Pollak_2018} and
$\dot x_b=p_b/m$ gives
$J_\Sigma(t)=\rho(x_b(t),t)g(t)/m$, where
$g(t)=\partial_xS(x_b(t),t)-p_b(t)$ and $g/m$ is the velocity relative to
the dividing surface.

If $X_b(t_1)\in\Gamma_\psi(t_1)$ at any finite time, invariance of the graph
and uniqueness of trajectories imply that $X_b$ lies on the graph for all
time, so $g\equiv0$ and $J_\Sigma\equiv0$. Otherwise $X_b$ is not a physical
characteristic, $g(t)\neq0$ for every finite $t$, and continuity fixes its
sign. Hence $J_\Sigma(t)$ is sign-definite and the probability current is
one-way. With the reactant-to-product orientation chosen as positive, the
instantaneous first-order decay rate is
$k(t)=J_\Sigma(t)/[N(t)-N_\infty]$, where
$N_\infty=\lim_{t\to\infty}N(t)$ is the residual reactant population; when
$N_\infty=0$, this reduces to $k(t)=J_\Sigma(t)/N(t)$
\cite{davis_bottlenecks_1985,gray_bottlenecks_1986,Dumont_Brumer_1986,hyland_toward_2014,craven_communication_2014,Feldmaier_Bardakcioglu_Reiff_Main_Hernandez_2019}.

For the Gaussian barrier-top model, the invariant manifolds separate the
physical characteristics into reactive and nonreactive sets. In the
displacement $\delta=(\delta_x,\delta_p)^T=X-X_b$, the physical graph is the
affine line $\delta_p=\beta\delta_x+g$, while the time slices
$W_t^u(X_b)$ and $W_t^s(X_b)$ are the lines
$\delta_p=s_u\delta_x$ and $\delta_p=s_s\delta_x$. If
$s_s<\beta<s_u$, the intersections
$Z_s(t)=W_t^s(X_b)\cap\Gamma_\psi(t)$ and
$Z_u(t)=W_t^u(X_b)\cap\Gamma_\psi(t)$ lie on opposite sides of
$\Sigma(t)=\{\delta_x=0\}$. They divide $\Gamma_\psi(t)$ into a middle
interval containing
$Z_\Sigma(t)=\Gamma_\psi(t)\cap\Sigma(t)$ and two exterior intervals.

Every physical characteristic can be written uniquely as
$\delta(t)=c_uX_u(t)+c_sX_s(t)$, with $c_u$ and $c_s$ constant along the
characteristic. Since $x_u$ and $x_s$ never vanish at finite time, we may
orient the modes so that $x_u,x_s>0$. Along $\Gamma_\psi(t)$, $c_u$
vanishes only at $Z_s$ and $c_s$ only at $Z_u$. At $Z_\Sigma$,
$0=c_ux_u+c_sx_s$, so the coefficients have opposite signs. By continuity,
they therefore have opposite signs throughout the middle interval and the
same sign in each exterior interval.

For a characteristic in the middle interval, the stable term dominates as
$t\to-\infty$ and the unstable term as $t\to+\infty$. Because their
coefficients have opposite signs, $\delta_x$ changes sign and must vanish.

It can do so only once: if $\delta_x(t_1)=\delta_x(t_2)=0$, multiplying
$\ddot\delta_x=\Omega^2(t)\delta_x$ by $\delta_x$ and integrating by parts
gives
\begin{equation}
\int_{t_1}^{t_2}
\left(\dot\delta_x^{\,2}+\Omega^2(t)\delta_x^2\right)\,dt=0,
\end{equation}
which is impossible for a nonzero displacement. Thus each middle-interval
characteristic crosses $\Sigma(t)$ exactly once. In the exterior intervals,
the two terms have the same sign, so $\delta_x$ never vanishes and no
crossing occurs.

Characteristics on the reactant side of $Z_\Sigma(t)$ cross in the future,
whereas those on the product side have already crossed. At any crossing,
$\delta_p=g(t)$ and hence $\dot\delta_x=g(t)/m$. Since $g$ has a fixed sign,
all crossings have the same orientation. The dividing surface is therefore
recrossing-free.

\begin{figure}
\centering
\includegraphics[width=0.65\linewidth]{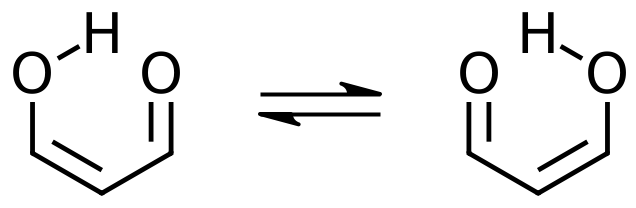}
\caption{Schematic of the \emph{cis}--\emph{cis} proton-transfer reaction in
malonaldehyde.}
\label{fig:ciscis}
\end{figure}

\emph{Worked example.---}
We apply the construction to a local model of \emph{cis}--\emph{cis} proton
transfer in malonaldehyde
\cite{cabral_roadmap_2024,ghosh_optimised_2015,ghosh_dynamics_2012}, shown
schematically in Fig.~\ref{fig:ciscis}. Along the reaction coordinate, we take
$V(x)=k_4x^4-k_2x^2$ and use the barrier-top approximation
$V(x)\approx-\tfrac12m\omega^2x^2$, with $\omega=0.00208741$ and $m=1836$
in atomic units from Ref.~\citenum{cabral_roadmap_2024}.

We use the inverted-harmonic-oscillator Gaussian wavepacket of
Ref.~\citenum{Heller_2018}. Let $\mathcal A_t$ denote its complex width
parameter and $q_t$ its center. Then
\begin{align}
a(t)&=\frac{4}{m}\bigl[\operatorname{Im}\mathcal A_t\bigr]^2,
&
f(t)&=-a(t)q_t.
\label{eq:func}
\end{align}
The expressions for $\mathcal A_t$ and $q_t$ are given in the SM. We take
$q_0=-1$, $p_0=3.6$, and
$\mathcal A_0=\tfrac12m\omega(1+i)$. Since $a(t)\ge0$,
$\Omega^2(t)=\omega^2+a(t)/m\ge\omega^2>0$. The SM shows that
$a(t)\to0$ as $|t|\to\infty$ and that
$f\in L^1(\mathbb R)\cap L^\infty(\mathbb R)$; hence
$A(t)\to A_\infty$. The conditions above are therefore satisfied,
establishing a unique bounded quantum transition-state trajectory.

\begin{figure}
\centering
\includegraphics[width=\linewidth]{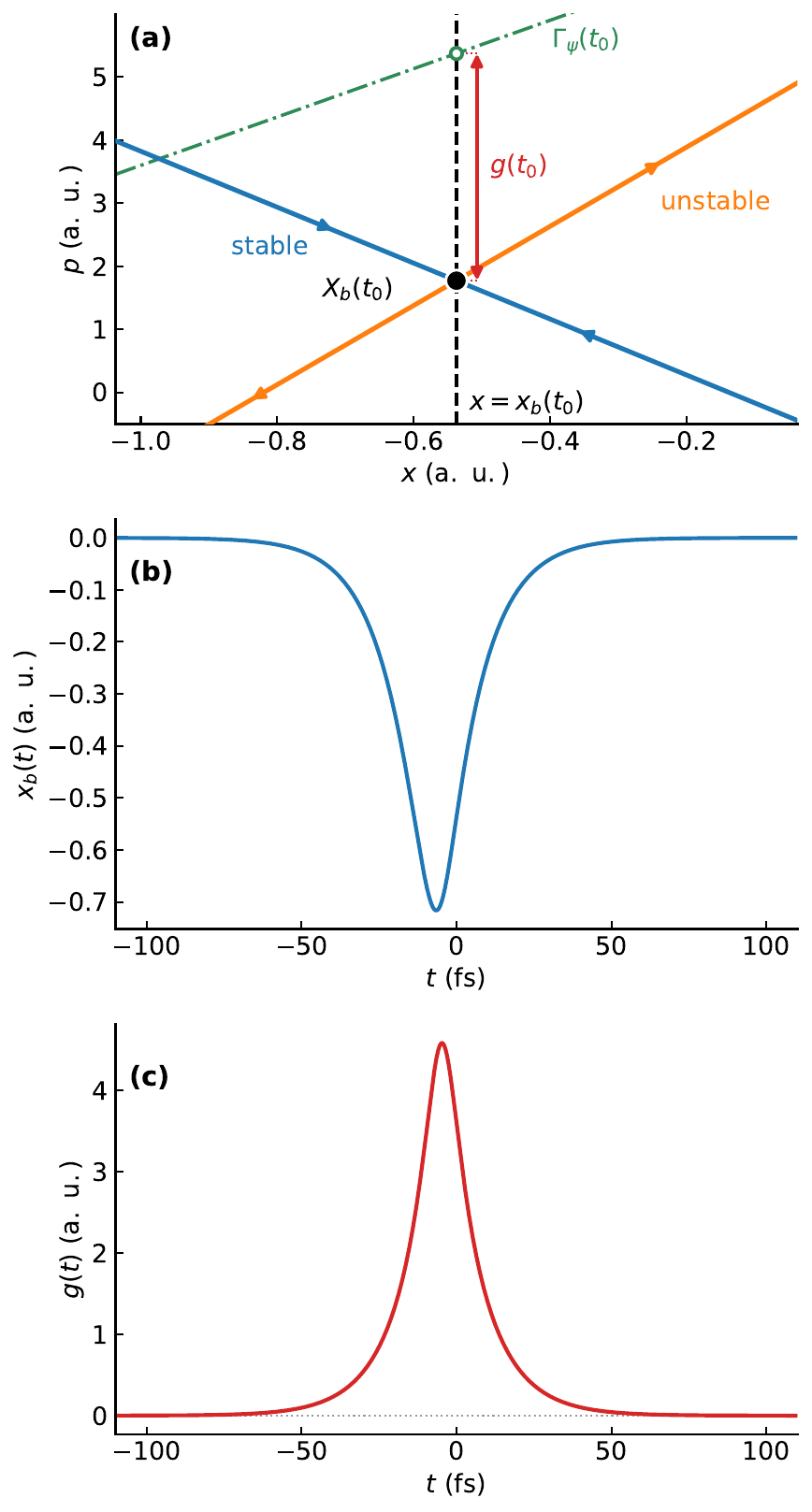}
\caption{Quantum transition-state structure for the malonaldehyde
proton-transfer barrier. (a)~At $t_0=0$, the time slices
$W_{t_0}^s(X_b)$ and $W_{t_0}^u(X_b)$ intersect at $X_b(t_0)$, whose
projection anchors the dividing surface $x=x_b(t_0)$. The arrow marks the
momentum gap $g(t_0)$ between $X_b(t_0)$ and $\Gamma_\psi(t_0)$.
(b)~The bounded coordinate $x_b(t)$ traces the moving dividing surface.
(c)~$g(t)$ remains positive, so the probability current through $\Sigma(t)$
is one-way.}
\label{fig:xb}
\end{figure}

To evaluate Eq.~\eqref{eq:bounded}, we obtain the homogeneous modes $X_s$ and
$X_u$ by initializing along the eigendirections of $A_\infty$ and integrating
backward from large positive time for $X_s$ and forward from large negative
time for $X_u$. For more general systems, the stable and unstable subspaces can be computed
numerically \cite{dieci_exponential_2010,dieci_detecting_2011}. With
$M(t)=\bigl(X_u(t)\ X_s(t)\bigr)$ and the constant
$W=\det M(t)$, the projector and propagator are
$P(t)=M(t)\operatorname{diag}(0,1)M^{-1}(t)$ and
$\Phi(t,\tau)=M(t)M^{-1}(\tau)$. Equation~\eqref{eq:bounded} then becomes
\begin{align}
X_b(t)
&=X_u(t)\int_t^\infty\frac{x_s(\tau)f(\tau)}{W}\,d\tau \notag\\
&\quad+X_s(t)\int_{-\infty}^t\frac{x_u(\tau)f(\tau)}{W}\,d\tau .
\label{eq:xpb}
\end{align}
The integration limits eliminate the growing homogeneous components, yielding
the bounded trajectory shown in Fig.~\ref{fig:xb}.

\begin{figure}
\centering
\includegraphics[width=\linewidth]{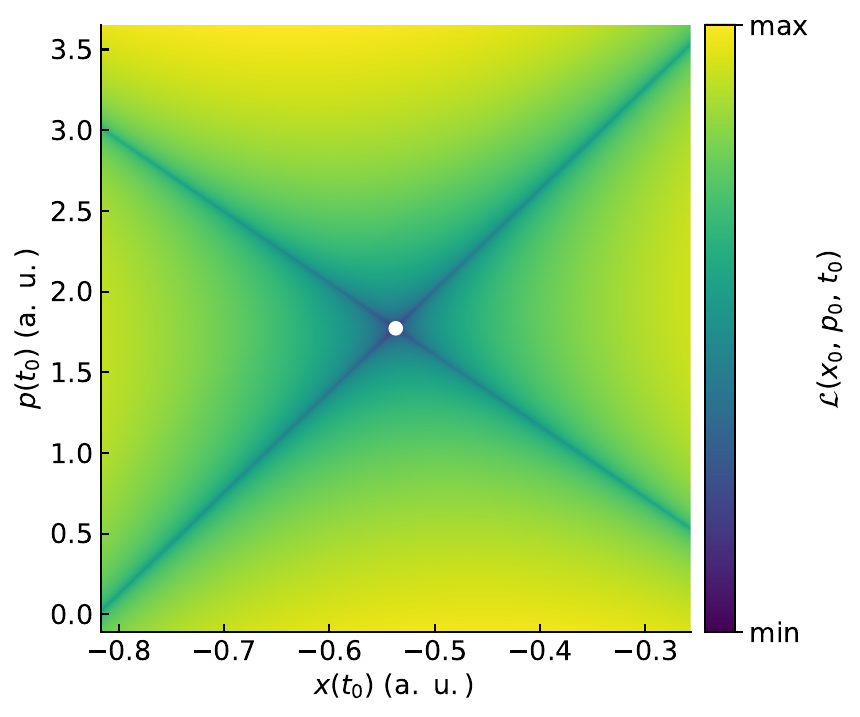}
\caption{Lagrangian descriptor for the malonaldehyde proton-transfer barrier
at $t_0=0$, with $\tau=4000\,\mathrm{a.u.}$ and $\alpha=0.2$. Its forward
and backward features trace $W_{t_0}^s(X_b)$ and $W_{t_0}^u(X_b)$; their
intersection (dot) is $X_b(t_0)$, which anchors the recrossing-free dividing
surface.}
\label{fig:ld}
\end{figure}

To visualize the phase-space geometry, we use the Lagrangian descriptor
\cite{Mancho_Wiggins_Curbelo_Mendoza_2013,Lopesino_etal_2017,Revuelta_Benito_Borondo_2021,junginger_transition_2016}
\begin{equation}
\mathcal L(x_0,p_0,t_0)
=\int_{t_0-\tau}^{t_0+\tau}|p(t)|^\alpha\,dt,
\label{eq:ld}
\end{equation}
a finite-time measure of momentum accumulation along a trajectory. We take
$\alpha=0.2$, with the same structure obtained for $0<\alpha\le1$. At fixed
$x_0$, minima of the forward and backward contributions over $p_0$
numerically identify $W_{t_0}^s(X_b)$ and $W_{t_0}^u(X_b)$, whose
intersection gives $X_b(t_0)$ (Fig.~\ref{fig:ld}). Source code used to generate Figs.~\ref{fig:xb} and \ref{fig:ld} is included
with the SM.

We have established a one-dimensional quantum transition state: the unique
bounded trajectory at which the stable and unstable manifold slices intersect.
Restricted to the monokinetic graph, these manifolds separate reactive from
nonreactive characteristics, with each reactive characteristic crossing the
anchored dividing surface exactly once and with the same orientation. The
resulting one-way probability current requires no simultaneous knowledge of
position and momentum direction; it follows from invariant geometry. In higher
dimensions, we anticipate that this trajectory generalizes to a time-dependent
normally hyperbolic center manifold, with additional center modes supplying its
internal directions, in analogy with classical periodic-orbit and NHIM
constructions
\cite{Pechukas_McLafferty_1973,pollak_transition_1978,wiggins_normally_1994,Ezra_Wiggins_2009,Collins_Ezra_Wiggins_2011,katsanikas_generalization_2021}.
The full multidimensional construction remains future work. Thus the
transition state is not destroyed by quantum mechanics; it becomes quantum.

\section{Acknowledgments}
\begin{acknowledgments}
The author thanks Florentino Borondo and Eitan Geva for their invaluable mentorship, and Stephen Wiggins for insightful comments on an earlier version of the manuscript. He also thanks his mother for her unwavering support and encouragement.

\end{acknowledgments}
\section*{Data Availability}

The numerical results shown in Figs.~\ref{fig:xb} and \ref{fig:ld} can be
reproduced using the source code and parameters provided with the Supplemental
Material~\cite{supplemental_material}.

\bibliography{quantum_transition_state}

@article{cabral_roadmap_2024,
  author        = {Cabral, Delmar G. A. and Khazaei, Pouya and Allen, Brandon C. and
                   Videla, Pablo E. and Sch{\"a}fer, Max and Corti{\~n}as, Rodrigo G. and
                   Carrillo De Albornoz, Alejandro Cros and Ch{\'a}vez-Carlos, Jorge and
                   Santos, Lea F. and Geva, Eitan and Batista, Victor S.},
  title         = {A roadmap for simulating chemical dynamics on a parametrically driven
                   bosonic quantum device},
  journal       = {J. Phys. Chem. Lett.},
  volume        = {15},
  pages         = {12042--12050},
  year          = {2024},
  doi           = {10.1021/acs.jpclett.4c02864},
}

@article{dieci_detecting_2011,
  author        = {Dieci, Luca and Elia, Cinzia and Van Vleck, Erik},
  title         = {Detecting exponential dichotomy on the real line: {SVD} and {QR}
                   algorithms},
  journal       = {BIT Numer. Math.},
  volume        = {51},
  pages         = {555--579},
  year          = {2011},
  doi           = {10.1007/s10543-010-0306-0},
}

@article{dieci_exponential_2010,
  author        = {Dieci, Luca and Elia, Cinzia and Van Vleck, Erik},
  title         = {Exponential dichotomy on the real line: {SVD} and {QR} methods},
  journal       = {J. Differ. Equ.},
  volume        = {248},
  pages         = {287--308},
  year          = {2010},
  doi           = {10.1016/j.jde.2009.07.004},
}

@book{coppel_dichotomies_1978,
  author        = {Coppel, W. A.},
  title         = {{Dichotomies in Stability Theory}},
  series        = {Lecture Notes in Mathematics},
  volume        = {629},
  publisher     = {Springer},
  address       = {Berlin},
  year          = {1978},
  doi           = {10.1007/BFb0067780},
}

@book{wiggins_normally_1994,
  author        = {Wiggins, Stephen},
  title         = {{Normally Hyperbolic Invariant Manifolds in Dynamical Systems}},
  series        = {Applied Mathematical Sciences},
  volume        = {105},
  publisher     = {Springer},
  address       = {New York},
  year          = {1994},
  doi           = {10.1007/978-1-4612-4312-0},
}

@book{barreira_stability_2008,
  author        = {Barreira, Luis and Valls, Claudia},
  title         = {{Stability of Nonautonomous Differential Equations}},
  series        = {Lecture Notes in Mathematics},
  volume        = {1926},
  publisher     = {Springer},
  address       = {Berlin},
  year          = {2008},
  doi           = {10.1007/978-3-540-74775-8},
}

@article{uzer_geometry_2002,
  author        = {Uzer, T. and Jaff{\'e}, Charles and Palaci{\'a}n, Jes{\'u}s and
                   Yanguas, Patricia and Wiggins, Stephen},
  title         = {The geometry of reaction dynamics},
  journal       = {Nonlinearity},
  volume        = {15},
  pages         = {957--992},
  year          = {2002},
  doi           = {10.1088/0951-7715/15/4/301},
}

@article{bartsch_transition_2005,
  author        = {Bartsch, Thomas and Hernandez, Rigoberto and Uzer, T.},
  title         = {Transition state in a noisy environment},
  journal       = {Phys. Rev. Lett.},
  volume        = {95},
  pages         = {058301},
  year          = {2005},
  doi           = {10.1103/PhysRevLett.95.058301},
}

@article{davis_bottlenecks_1985,
  author        = {Davis, Michael J.},
  title         = {Bottlenecks to intramolecular energy transfer and the calculation of
                   relaxation rates},
  journal       = {J. Chem. Phys.},
  volume        = {83},
  pages         = {1016--1031},
  year          = {1985},
  doi           = {10.1063/1.449465},
}

@article{gray_bottlenecks_1986,
  author        = {Gray, Stephen K. and Rice, Stuart A. and Davis, Michael J},
  title         = {Bottlenecks to unimolecular reactions and an alternative form for
                   classical {RRKM} theory},
  journal       = {J. Phys. Chem.},
  volume        = {90},
  pages         = {3470--3482},
  year          = {1986},
  doi           = {10.1021/j100407a005},
}

@book{Norsen_2017,
  author        = {Norsen, Travis},
  title         = {{Foundations of Quantum Mechanics: An Exploration of the Physical
                   Meaning of Quantum Theory}},
  series        = {Undergraduate Lecture Notes in Physics},
  publisher     = {Springer},
  address       = {Cham},
  year          = {2017},
  doi           = {10.1007/978-3-319-65867-4},
}

@article{ghosh_dynamics_2012,
  author        = {Ghosh, S and Bhattacharyya, S P},
  title         = {Dynamics of atom tunnelling in a symmetric double well coupled to an
                   asymmetric double well: {The} case of malonaldehyde},
  journal       = {J. Chem. Sci.},
  volume        = {124},
  pages         = {13--19},
  year          = {2012},
  doi           = {10.1007/s12039-011-0205-1},
}

@article{ghosh_optimised_2015,
  author        = {Ghosh, Subhasree and Talukder, Srijeeta and Sen, Shrabani and
                   Chaudhury, Pinaki},
  title         = {Optimised polychromatic field-mediated suppression of {H}-atom
                   tunnelling in a coupled symmetric double well: two-dimensional
                   malonaldehyde model},
  journal       = {Mol. Phys.},
  volume        = {113},
  pages         = {3826--3838},
  year          = {2015},
  doi           = {10.1080/00268976.2015.1068393},
}

@article{bohm_suggested_1952,
  author        = {Bohm, David},
  title         = {A suggested interpretation of the quantum theory in terms of ``hidden''
                   variables. {I}},
  journal       = {Phys. Rev.},
  volume        = {85},
  pages         = {166--179},
  year          = {1952},
  doi           = {10.1103/PhysRev.85.166},
}

@book{Durr_Teufel_2009,
  author        = {D{\"u}rr, Detlef and Teufel, Stefan},
  title         = {{Bohmian Mechanics: The Physics and Mathematics of Quantum Theory}},
  publisher     = {Springer},
  address       = {Berlin},
  year          = {2009},
  doi           = {10.1007/b99978},
}

@article{Revuelta_Benito_Borondo_2021,
  author        = {Revuelta, F. and Benito, R. M. and Borondo, F.},
  title         = {Identification of the invariant manifolds of the {LiCN} molecule using
                   {Lagrangian} descriptors},
  journal       = {Phys. Rev. E},
  volume        = {104},
  pages         = {044210},
  year          = {2021},
  doi           = {10.1103/PhysRevE.104.044210},
}

@article{junginger_transition_2016,
  author        = {Junginger, Andrej and Craven, Galen T. and Bartsch, Thomas and
                   Revuelta, F. and Borondo, F. and Benito, R. M. and Hernandez, Rigoberto},
  title         = {Transition state geometry of driven chemical reactions on
                   time-dependent double-well potentials},
  journal       = {Phys. Chem. Chem. Phys.},
  volume        = {18},
  pages         = {30270--30281},
  year          = {2016},
  doi           = {10.1039/C6CP02519F},
}

@article{wiggins_role_2016,
  author        = {Wiggins, Stephen},
  title         = {The role of normally hyperbolic invariant manifolds ({NHIMs}) in the
                   context of the phase space setting for chemical reaction dynamics},
  journal       = {Regul. Chaotic Dyn.},
  volume        = {21},
  pages         = {621--638},
  year          = {2016},
  doi           = {10.1134/S1560354716060034},
}

@book{Heller_2018,
  author        = {Heller, Eric J.},
  title         = {{The Semiclassical Way to Dynamics and Spectroscopy}},
  publisher     = {Princeton University Press},
  address       = {Princeton, NJ},
  year          = {2018},
}

@article{pollak_quantum_1993,
  author        = {Pollak, Eli and Proselkov, Dmitry},
  title         = {Quantum variational transition state theory revisited},
  journal       = {Chem. Phys.},
  volume        = {170},
  pages         = {265--273},
  year          = {1993},
  doi           = {10.1016/0301-0104(93)85113-M},
}

@article{hele_derivation_2013,
  author        = {Hele, Timothy J. H. and Althorpe, Stuart C.},
  title         = {Derivation of a true ($t \to 0^+$) quantum transition-state theory.
                   {I}. Uniqueness and equivalence to ring-polymer molecular dynamics
                   transition-state theory},
  journal       = {J. Chem. Phys.},
  volume        = {138},
  pages         = {084108},
  year          = {2013},
  doi           = {10.1063/1.4792697},
}

@article{bohm_suggested_1952-1,
  author        = {Bohm, David},
  title         = {A suggested interpretation of the quantum theory in terms of ``hidden''
                   variables. {II}},
  journal       = {Phys. Rev.},
  volume        = {85},
  pages         = {180--193},
  year          = {1952},
  doi           = {10.1103/PhysRev.85.180},
}

@article{hyland_toward_2014,
  author        = {Hyland, Brittany L. and Martens, Craig C.},
  title         = {Toward a quantum trajectory-based rate theory},
  journal       = {Theor. Chem. Acc.},
  volume        = {133},
  pages         = {1536},
  year          = {2014},
  doi           = {10.1007/s00214-014-1536-z},
}

@incollection{rice_time-dependent_2008,
  author        = {Bartsch, Thomas and Moix, Jeremy M. and Hernandez, Rigoberto and Kawai,
                   Shinnosuke and Uzer, T.},
  title         = {Time-dependent transition state theory},
  booktitle     = {Advances in Chemical Physics},
  editor        = {Rice, Stuart A.},
  volume        = {140},
  pages         = {191--238},
  publisher     = {Wiley},
  address       = {Hoboken, NJ},
  year          = {2008},
  doi           = {10.1002/9780470371572.ch4},
}

@article{craven_communication_2014,
  author        = {Craven, Galen T. and Bartsch, Thomas and Hernandez, Rigoberto},
  title         = {Communication: transition state trajectory stability determines barrier
                   crossing rates in chemical reactions induced by time-dependent
                   oscillating fields},
  journal       = {J. Chem. Phys.},
  volume        = {141},
  pages         = {041106},
  year          = {2014},
  doi           = {10.1063/1.4891471},
}

@article{ju_roughness_2001,
  author        = {Ju, Ning and Wiggins, Stephen},
  title         = {On roughness of exponential dichotomy},
  journal       = {J. Math. Anal. Appl.},
  volume        = {262},
  pages         = {39--49},
  year          = {2001},
  doi           = {10.1006/jmaa.2001.7496},
}

@article{Truhlar_Garrett_Klippenstein_1996,
  author        = {Truhlar, Donald G and Garrett, Bruce C and Klippenstein, Stephen J},
  title         = {Current status of transition-state theory},
  journal       = {J. Phys. Chem.},
  volume        = {100},
  pages         = {12771--12800},
  year          = {1996},
  doi           = {10.1021/jp953748q},
}

@article{miller_quantum_1983,
  author        = {Miller, William H. and Schwartz, Steven D. and Tromp, John W.},
  title         = {Quantum mechanical rate constants for bimolecular reactions},
  journal       = {J. Chem. Phys.},
  volume        = {79},
  pages         = {4889--4898},
  year          = {1983},
  doi           = {10.1063/1.445581},
}

@article{hirschfelder_quantum-mechanical_1939,
  author        = {Hirschfelder, J. O. and Wigner, E.},
  title         = {Some quantum-mechanical considerations in the theory of reactions
                   involving an activation energy},
  journal       = {J. Chem. Phys.},
  volume        = {7},
  pages         = {616--628},
  year          = {1939},
  doi           = {10.1063/1.1750500},
}

@article{yamamoto_quantum_1960,
  author        = {Yamamoto, Tsunenobu},
  title         = {Quantum statistical mechanical theory of the rate of exchange chemical
                   reactions in the gas phase},
  journal       = {J. Chem. Phys.},
  volume        = {33},
  pages         = {281--289},
  year          = {1960},
  doi           = {10.1063/1.1731099},
}

@article{predescu_optimal_2005,
  author        = {Predescu, Cristian and Miller, William H.},
  title         = {Optimal choice of dividing surface for the computation of quantum
                   reaction rates},
  journal       = {J. Phys. Chem. B},
  volume        = {109},
  pages         = {6491--6499},
  year          = {2005},
  doi           = {10.1021/jp040593q},
}

@article{small_quantifying_2006,
  author        = {Small, Michael S. and Predescu, Cristian and Miller, William H.},
  title         = {Quantifying the extent of recrossing flux for quantum systems},
  journal       = {Chem. Phys.},
  volume        = {322},
  pages         = {151--159},
  year          = {2006},
  doi           = {10.1016/j.chemphys.2005.07.036},
}

@article{pollak_transition_1978,
  author        = {Pollak, Eli and Pechukas, Philip},
  title         = {Transition states, trapped trajectories, and classical bound states
                   embedded in the continuum},
  journal       = {J. Chem. Phys.},
  volume        = {69},
  pages         = {1218--1226},
  year          = {1978},
  doi           = {10.1063/1.436658},
}

@article{katsanikas_generalization_2021,
  author        = {Katsanikas, Matthaios and Wiggins, Stephen},
  title         = {The generalization of the periodic-orbit dividing surface in
                   {Hamiltonian} systems with three or more degrees of freedom---{I}},
  journal       = {Int. J. Bifurcation Chaos},
  volume        = {31},
  pages         = {2130028},
  year          = {2021},
  doi           = {10.1142/S0218127421300287},
}

@article{miller_beyond_1993,
  author        = {Miller, William H.},
  title         = {Beyond transition-state theory: a rigorous quantum theory of chemical
                   reaction rates},
  journal       = {Acc. Chem. Res.},
  volume        = {26},
  pages         = {174--181},
  year          = {1993},
  doi           = {10.1021/ar00028a007},
}

@article{miller_quantum_1974,
  author        = {Miller, William H.},
  title         = {Quantum mechanical transition state theory and a new semiclassical
                   model for reaction rate constants},
  journal       = {J. Chem. Phys.},
  volume        = {61},
  pages         = {1823--1834},
  year          = {1974},
  doi           = {10.1063/1.1682181},
}

@article{pollak_new_1998,
  author        = {Pollak, Eli and Liao, Jie-Lou},
  title         = {A new quantum transition state theory},
  journal       = {J. Chem. Phys.},
  volume        = {108},
  pages         = {2733--2743},
  year          = {1998},
  doi           = {10.1063/1.475665},
}

@article{Eyring_1938_1,
  author        = {Eyring, Henry},
  title         = {The theory of absolute reaction rates},
  journal       = {Trans. Faraday Soc.},
  volume        = {34},
  pages         = {41--48},
  year          = {1938},
  doi           = {10.1039/tf9383400041},
}

@article{Eyring_1938,
  author        = {Eyring, Henry},
  title         = {The calculation of activation energies},
  journal       = {Trans. Faraday Soc.},
  volume        = {34},
  pages         = {3--11},
  year          = {1938},
  doi           = {10.1039/tf9383400003},
}

@article{wiggins_geometric_2026,
  author        = {Wiggins, Stephen},
  title         = {Geometric diagnostics of scrambling-related sensitivity in a {Bohmian}
                   preparation space},
  journal       = {J. Chem. Phys.},
  volume        = {164},
  pages         = {244108},
  year          = {2026},
  doi           = {10.1063/5.0339097},
}

@article{Wigner_1938,
  author        = {Wigner, E.},
  title         = {The transition state method},
  journal       = {Trans. Faraday Soc.},
  volume        = {34},
  pages         = {29--41},
  year          = {1938},
  doi           = {10.1039/tf9383400029},
}

@article{Evans_Polanyi_1935,
  author        = {Evans, M. G. and Polanyi, M.},
  title         = {Some applications of the transition state method to the calculation of
                   reaction velocities, especially in solution},
  journal       = {Trans. Faraday Soc.},
  volume        = {31},
  pages         = {875--894},
  year          = {1935},
  doi           = {10.1039/tf9353100875},
}

@article{miller_spiers_1998,
  author        = {Miller, William H.},
  title         = {Spiers Memorial Lecture: Quantum and semiclassical theory of chemical
                   reaction rates},
  journal       = {Faraday Discuss.},
  volume        = {110},
  pages         = {1--21},
  year          = {1998},
  doi           = {10.1039/a805196h},
}

@article{Pechukas_McLafferty_1973,
  author        = {Pechukas, Philip and McLafferty, Francis J.},
  title         = {On transition-state theory and the classical mechanics of collinear
                   collisions},
  journal       = {J. Chem. Phys.},
  volume        = {58},
  pages         = {1622--1625},
  year          = {1973},
  doi           = {10.1063/1.1679404},
}

@article{Voth_Chandler_Miller_1989,
  author        = {Voth, Gregory A. and Chandler, David and Miller, William H.},
  title         = {Rigorous formulation of quantum transition state theory and its
                   dynamical corrections},
  journal       = {J. Chem. Phys.},
  volume        = {91},
  pages         = {7749--7760},
  year          = {1989},
  doi           = {10.1063/1.457242},
}

@article{wigner_crossing_1932,
  author        = {Wigner, E.},
  title         = {{\"U}ber das {\"U}berschreiten von Potentialschwellen bei chemischen
                   Reaktionen},
  journal       = {Z. Phys. Chem. B},
  volume        = {19},
  pages         = {203--216},
  year          = {1932},
}

@article{Waalkens_Schubert_Wiggins_2008,
  author        = {Waalkens, Holger and Schubert, Roman and Wiggins, Stephen},
  title         = {{Wigner}'s dynamical transition state theory in phase space: classical
                   and quantum},
  journal       = {Nonlinearity},
  volume        = {21},
  pages         = {R1--R118},
  year          = {2008},
  doi           = {10.1088/0951-7715/21/1/R01},
}

@article{Geva_Shi_Voth_2001,
  author        = {Geva, Eitan and Shi, Qiang and Voth, Gregory A.},
  title         = {Quantum-mechanical reaction rate constants from centroid molecular
                   dynamics simulations},
  journal       = {J. Chem. Phys.},
  volume        = {115},
  pages         = {9209--9222},
  year          = {2001},
  doi           = {10.1063/1.1412870},
}

@article{McLafferty_Pechukas_1974,
  author        = {McLafferty, Francis J. and Pechukas, Philip},
  title         = {Quantum transition state theory},
  journal       = {Chem. Phys. Lett.},
  volume        = {27},
  pages         = {511--514},
  year          = {1974},
  doi           = {10.1016/0009-2614(74)80293-9},
}

@article{pollak_reaction_2005,
  author        = {Pollak, Eli and Talkner, Peter},
  title         = {Reaction rate theory: what it was, where is it today, and where is it
                   going?},
  journal       = {Chaos},
  volume        = {15},
  pages         = {026116},
  year          = {2005},
  doi           = {10.1063/1.1858782},
}

@article{Richardson_Althorpe_2009,
  author        = {Richardson, Jeremy O. and Althorpe, Stuart C.},
  title         = {Ring-polymer molecular dynamics rate theory in the deep-tunneling
                   regime: connection with semiclassical instanton theory},
  journal       = {J. Chem. Phys.},
  volume        = {131},
  pages         = {214106},
  year          = {2009},
  doi           = {10.1063/1.3267318},
}

@article{Richardson_2016,
  author        = {Richardson, Jeremy O.},
  title         = {Derivation of instanton rate theory from first principles},
  journal       = {J. Chem. Phys.},
  volume        = {144},
  pages         = {114106},
  year          = {2016},
  doi           = {10.1063/1.4943866},
}

@article{Craig_Manolopoulos_2005,
  author        = {Craig, Ian R. and Manolopoulos, David E.},
  title         = {Chemical reaction rates from ring polymer molecular dynamics},
  journal       = {J. Chem. Phys.},
  volume        = {122},
  pages         = {084106},
  year          = {2005},
  doi           = {10.1063/1.1850093},
}

@article{Topaler_Makri_1993,
  author        = {Topaler, Maria and Makri, Nancy},
  title         = {Quasi-adiabatic propagator path integral methods: exact quantum rate
                   constants for condensed-phase reactions},
  journal       = {Chem. Phys. Lett.},
  volume        = {210},
  pages         = {285--293},
  year          = {1993},
  doi           = {10.1016/0009-2614(93)89135-5},
}

@article{Topaler_Makri_1994,
  author        = {Topaler, Maria and Makri, Nancy},
  title         = {Quantum rates for a double well coupled to a dissipative bath: accurate
                   path integral results and comparison with approximate theories},
  journal       = {J. Chem. Phys.},
  volume        = {101},
  pages         = {7500--7519},
  year          = {1994},
  doi           = {10.1063/1.468244},
}

@article{Donoso_Martens_1998,
  author        = {Donoso, Arnaldo and Martens, Craig C},
  title         = {Simulation of coherent nonadiabatic dynamics using classical
                   trajectories},
  journal       = {J. Phys. Chem. A},
  volume        = {102},
  pages         = {4291--4300},
  year          = {1998},
  doi           = {10.1021/jp980219o},
}

@book{Wyatt_Trahan_2005,
  author        = {Wyatt, Robert E. and Trahan, Corey J.},
  title         = {{Quantum Dynamics with Trajectories: Introduction to Quantum
                   Hydrodynamics}},
  series        = {Interdisciplinary Applied Mathematics},
  volume        = {28},
  publisher     = {Springer},
  address       = {New York},
  year          = {2005},
  doi           = {10.1007/0-387-28145-2},
}

@article{Garashchuk_Rassolov_2003,
  author        = {Garashchuk, Sophya and Rassolov, Vitaly A.},
  title         = {Quantum dynamics with {Bohmian} trajectories: energy conserving
                   approximation to the quantum potential},
  journal       = {Chem. Phys. Lett.},
  volume        = {376},
  pages         = {358--363},
  year          = {2003},
  doi           = {10.1016/S0009-2614(03)01008-X},
}

@article{donoso_quantum_2001,
  author        = {Donoso, Arnaldo and Martens, Craig C.},
  title         = {Quantum tunneling using entangled classical trajectories},
  journal       = {Phys. Rev. Lett.},
  volume        = {87},
  pages         = {223202},
  year          = {2001},
  doi           = {10.1103/PhysRevLett.87.223202},
}

@article{garashchuk_semiclassical_2002,
  author        = {Garashchuk, Sophya and Rassolov, Vitaly A.},
  title         = {Semiclassical dynamics based on quantum trajectories},
  journal       = {Chem. Phys. Lett.},
  volume        = {364},
  pages         = {562--567},
  year          = {2002},
  doi           = {10.1016/S0009-2614(02)01378-7},
}

@article{maddox_quantum_2002,
  author        = {Maddox, Jeremy B. and Bittner, Eric R.},
  title         = {Quantum dissipation in unbounded systems},
  journal       = {Phys. Rev. E},
  volume        = {65},
  pages         = {026143},
  year          = {2002},
  doi           = {10.1103/PhysRevE.65.026143},
}

@article{maddox_quantum_2001,
  author        = {Maddox, Jeremy B. and Bittner, Eric R.},
  title         = {Quantum relaxation dynamics using {Bohmian} trajectories},
  journal       = {J. Chem. Phys.},
  volume        = {115},
  pages         = {6309--6316},
  year          = {2001},
  doi           = {10.1063/1.1394747},
}

@article{Sanz_Borondo_2007,
  author        = {Sanz, A. S. and Borondo, F.},
  title         = {A quantum trajectory description of decoherence},
  journal       = {Eur. Phys. J. D},
  volume        = {44},
  pages         = {319--326},
  year          = {2007},
  doi           = {10.1140/epjd/e2007-00191-8},
}

@article{Goldfarb_Degani_Tannor_2006,
  author        = {Goldfarb, Yair and Degani, Ilan and Tannor, David J.},
  title         = {{Bohmian} mechanics with complex action: a new trajectory-based
                   formulation of quantum mechanics},
  journal       = {J. Chem. Phys.},
  volume        = {125},
  pages         = {231103},
  year          = {2006},
  doi           = {10.1063/1.2400851},
}

@article{liu_monte_2004,
  author        = {Liu, Jian and Makri, Nancy},
  title         = {{Monte Carlo} {Bohmian} dynamics from trajectory stability properties},
  journal       = {J. Phys. Chem. A},
  volume        = {108},
  pages         = {5408--5416},
  year          = {2004},
  doi           = {10.1021/jp040149n},
}

@article{liu_bohms_2005,
  author        = {Liu, Jian and Makri, Nancy},
  title         = {{Bohm}'s formulation in imaginary time: estimation of energy eigenvalues},
  journal       = {Mol. Phys.},
  volume        = {103},
  pages         = {1083--1090},
  year          = {2005},
  doi           = {10.1080/00268970512331339387},
}

@article{sanz_contextuality_2009,
  author        = {Sanz, A. S. and Borondo, F.},
  title         = {Contextuality, decoherence and quantum trajectories},
  journal       = {Chem. Phys. Lett.},
  volume        = {478},
  pages         = {301--306},
  year          = {2009},
  doi           = {10.1016/j.cplett.2009.07.061},
}

@article{Collins_Ezra_Wiggins_2011,
  author        = {Collins, Peter and Ezra, Gregory S. and Wiggins, Stephen},
  title         = {Index k saddles and dividing surfaces in phase space with applications
                   to isomerization dynamics},
  journal       = {J. Chem. Phys.},
  volume        = {134},
  pages         = {244105},
  year          = {2011},
  doi           = {10.1063/1.3602465},
}

@article{Dumont_Brumer_1986,
  author        = {Dumont, Randall S and Brumer, Paul},
  title         = {Dynamical theory of statistical unimolecular decay},
  journal       = {J. Phys. Chem.},
  volume        = {90},
  pages         = {3509--3516},
  year          = {1986},
  doi           = {10.1021/j100407a012},
}

@article{Ezra_Wiggins_2009,
  author        = {Ezra, Gregory S and Wiggins, Stephen},
  title         = {Phase-space geometry and reaction dynamics near index 2 saddles},
  journal       = {J. Phys. A: Math. Theor.},
  volume        = {42},
  pages         = {205101},
  year          = {2009},
  doi           = {10.1088/1751-8113/42/20/205101},
}

@article{Cohen_Pollak_2018,
  author        = {Cohen, Eliahu and Pollak, Eli},
  title         = {Determination of weak values of quantum operators using only strong
                   measurements},
  journal       = {Phys. Rev. A},
  volume        = {98},
  pages         = {042112},
  year          = {2018},
  doi           = {10.1103/PhysRevA.98.042112},
}

@article{Borondo_Luque_Villanueva_Wisniacki_2009,
  author        = {Borondo, F and Luque, A and Villanueva, J and Wisniacki, D A},
  title         = {A dynamical systems approach to {Bohmian} trajectories in a {2D} harmonic
                   oscillator},
  journal       = {J. Phys. A: Math. Theor.},
  volume        = {42},
  pages         = {495103},
  year          = {2009},
  doi           = {10.1088/1751-8113/42/49/495103},
}

@article{Wisniacki_Pujals_Borondo_2006,
  author        = {Wisniacki, D. A and Pujals, E. R and Borondo, F},
  title         = {Vortex interaction, chaos and quantum probabilities},
  journal       = {Europhys. Lett.},
  volume        = {73},
  pages         = {671--676},
  year          = {2006},
  doi           = {10.1209/epl/i2005-10467-5},
}

@article{Wisniacki_Pujals_2005,
  author        = {Wisniacki, D. A and Pujals, E. R},
  title         = {Motion of vortices implies chaos in {Bohmian} mechanics},
  journal       = {Europhys. Lett.},
  volume        = {71},
  pages         = {159--165},
  year          = {2005},
  doi           = {10.1209/epl/i2005-10085-3},
}

@article{Martens_2002,
  author        = {Martens, Craig C.},
  title         = {Qualitative dynamics of generalized {Langevin} equations and the theory
                   of chemical reaction rates},
  journal       = {J. Chem. Phys.},
  volume        = {116},
  pages         = {2516--2528},
  year          = {2002},
  doi           = {10.1063/1.1436116},
}

@article{Mancho_Wiggins_Curbelo_Mendoza_2013,
  author        = {Mancho, Ana M. and Wiggins, Stephen and Curbelo, Jezabel and Mendoza,
                   Carolina},
  title         = {{Lagrangian} descriptors: a method for revealing phase-space structures
                   of general time-dependent dynamical systems},
  journal       = {Commun. Nonlinear Sci. Numer. Simul.},
  volume        = {18},
  pages         = {3530--3557},
  year          = {2013},
  doi           = {10.1016/j.cnsns.2013.05.002},
}

@article{Lopesino_etal_2017,
  author        = {Lopesino, C. and Balibrea-Iniesta, F. and Garc{\'i}a-Garrido, V. J. and
                   Wiggins, S. and Mancho, A. M.},
  title         = {A theoretical framework for {Lagrangian} descriptors},
  journal       = {Int. J. Bifurcation Chaos},
  volume        = {27},
  pages         = {1730001},
  year          = {2017},
  doi           = {10.1142/S0218127417300014},
}

@article{Jang_Voth_2016,
  author        = {Jang, Seogjoo and Voth, Gregory A.},
  title         = {Can quantum transition-state theory be defined as an exact {$t \to
                   0^+$} limit?},
  journal       = {J. Chem. Phys.},
  volume        = {144},
  pages         = {084110},
  year          = {2016},
  doi           = {10.1063/1.4942482},
}

@article{madelung_quantentheorie_1927,
  author        = {Madelung, Erwin},
  title         = {Quantentheorie in hydrodynamischer {Form}},
  journal       = {Z. Phys.},
  volume        = {40},
  pages         = {322--326},
  year          = {1927},
  doi           = {10.1007/BF01400372},
}

@article{Feldmaier_Bardakcioglu_Reiff_Main_Hernandez_2019,
  author        = {Feldmaier, Matthias and Bardakcioglu, Robin and Reiff, Johannes and
                   Main, J{\"o}rg and Hernandez, Rigoberto},
  title         = {Phase-space resolved rates in driven multidimensional chemical
                   reactions},
  journal       = {J. Chem. Phys.},
  volume        = {151},
  pages         = {244108},
  year          = {2019},
  doi           = {10.1063/1.5127539},
}

@article{Durr_Goldstein_Zanghi_1992,
  author        = {D{\"u}rr, Detlef and Goldstein, Sheldon and Zangh{\`i}, Nino},
  title         = {Quantum equilibrium and the origin of absolute uncertainty},
  journal       = {J. Stat. Phys.},
  volume        = {67},
  pages         = {843--907},
  year          = {1992},
  doi           = {10.1007/BF01049004},
}

@article{Daumer_Durr_Goldstein_Zanghi_1997,
  author        = {Daumer, Martin and D{\"u}rr, Detlef and Goldstein, Sheldon and
                   Zangh{\`i}, Nino},
  title         = {On the quantum probability flux through surfaces},
  journal       = {J. Stat. Phys.},
  volume        = {88},
  pages         = {967--977},
  year          = {1997},
  doi           = {10.1023/B:JOSS.0000015181.86864.fb},
}

@book{Bricmont_2016,
  author        = {Bricmont, Jean},
  title         = {{Making Sense of Quantum Mechanics}},
  publisher     = {Springer},
  address       = {Cham},
  year          = {2016},
  doi           = {10.1007/978-3-319-25889-8},
}

@book{Holland_1993,
  author        = {Holland, Peter R.},
  title         = {{The Quantum Theory of Motion: An Account of the
                   de Broglie--Bohm Causal Interpretation of Quantum Mechanics}},
  publisher     = {Cambridge University Press},
  address       = {Cambridge},
  year          = {1993},
  doi           = {10.1017/CBO9780511622687},
}

@article{deLeon_Sardon_2017,
  author        = {de Le{\'o}n, M. and Sard{\'o}n, C.},
  title         = {Cosymplectic and contact structures for time-dependent and dissipative
                   {Hamiltonian} systems},
  journal       = {J. Phys. A: Math. Theor.},
  volume        = {50},
  pages         = {255205},
  year          = {2017},
  doi           = {10.1088/1751-8121/aa711d},
}

@article{RomanRoy_2021,
  author        = {Rom{\'a}n-Roy, Narciso},
  title         = {An overview of the Hamilton--Jacobi theory: the classical and
                   geometrical approaches and some extensions and applications},
  journal       = {Mathematics},
  volume        = {9},
  pages         = {85},
  year          = {2021},
  doi           = {10.3390/math9010085},
}

@article{Eyring_1935,
  author  = {Eyring, Henry},
  title   = {The activated complex in chemical reactions},
  journal = {J. Chem. Phys.},
  volume  = {3},
  number  = {2},
  pages   = {107--115},
  year    = {1935},
  doi     = {10.1063/1.1749604},
}

@article{Schubert_Waalkens_Wiggins_2006,
  author  = {Schubert, Roman and Waalkens, Holger and Wiggins, Stephen},
  title   = {Efficient computation of transition-state resonances and reaction
             rates from a quantum normal form},
  journal = {Phys. Rev. Lett.},
  volume  = {96},
  pages   = {218302},
  year    = {2006},
  doi     = {10.1103/PhysRevLett.96.218302},
}

@article{Pelzer_Wigner_1932,
  author  = {Pelzer, H. and Wigner, E.},
  title   = {{\"U}ber die {Geschwindigkeitskonstante} von
             {Austauschreaktionen}},
  journal = {Z. Phys. Chem. B},
  volume  = {15},
  pages   = {445--471},
  year    = {1932},
  doi     = {10.1515/zpch-1932-1539},
}

@article{Eyring_Polanyi_1931,
  author  = {Eyring, Henry and Polanyi, Michael},
  title   = {{\"U}ber einfache {Gasreaktionen}},
  journal = {Z. Phys. Chem. B},
  volume  = {12},
  pages   = {279--311},
  year    = {1931},
}

@article{Jang_Voth_2000,
  author        = {Jang, Seogjoo and Voth, Gregory A.},
  title         = {A relationship between centroid dynamics and path integral quantum
                   transition state theory},
  journal       = {J. Chem. Phys.},
  volume        = {112},
  pages         = {8747--8757},
  year          = {2000},
  doi           = {10.1063/1.481490},
}

@article{Jang_Voth_2017,
  author        = {Jang, Seogjoo and Voth, Gregory A.},
  title         = {Non-uniqueness of quantum transition state theory and general dividing
                   surfaces in the path integral space},
  journal       = {J. Chem. Phys.},
  volume        = {146},
  pages         = {174106},
  year          = {2017},
  doi           = {10.1063/1.4982053},
}

@article{Shi_Geva_2002,
  author        = {Shi, Qiang and Geva, Eitan},
  title         = {Centroid-based methods for calculating quantum reaction rate constants:
                   centroid sampling versus centroid dynamics},
  journal       = {J. Chem. Phys.},
  volume        = {116},
  pages         = {3223--3233},
  year          = {2002},
  doi           = {10.1063/1.1445120},
}

@misc{supplemental_material,
  note = {See Supplemental Material at [URL will be inserted by publisher]
          for background on the dynamical-systems and quantum-trajectory
          framework, derivations for the Gaussian barrier-top model, and
          source code used to generate the figures.},
}

\end{document}


\title{Supplemental Material for ``The Quantum Transition State''}
\author{Pouya Khazaei}
\email{pkhazaei@umich.edu}
\affiliation{Department of Chemistry, University of Michigan, Ann Arbor, MI 48109, U.S.A.}
\affiliation{Chemical Physics Theory Group, Department of Chemistry, University of Toronto, Toronto, Ontario M5S 3H6, Canada}
\affiliation{Department of Physical and Environmental Sciences, University of Toronto Scarborough, Toronto, Ontario M1C 1A4, Canada}
\date{\today}
\maketitle
\onecolumngrid

This Supplemental Material introduces the dynamical-systems and quantum-
trajectory ideas needed in the Letter. Section~\ref{sec:background} explains
flows, invariant manifolds, exponential dichotomies, transition-state
geometry, Lagrangian descriptors, and quantum trajectories.
Section~\ref{sec:gaussian} gives the details of the Gaussian barrier-top model
and verifies the properties used in the Letter.

\section{Background}
\label{sec:background}

\subsection{Trajectories, flows, and invariant manifolds}

A dynamical system describes how a state changes with time:
\begin{equation}
\dot X=\mathcal F(X,t).
\label{sm:ode}
\end{equation}
In the Letter, $X=(x,p)^T$ is a point in the two-dimensional phase space
$\mathcal P\simeq\R^2$. A \emph{trajectory} is the path $X(t)$ generated by
one initial condition. The \emph{flow} evolves all initial conditions: if the
state is $X_0$ at time $t_0$, then
\begin{equation}
X(t)=\varphi_{t,t_0}(X_0).
\label{sm:flow}
\end{equation}
When $\mathcal F$ depends explicitly on time, the system is
\emph{nonautonomous}, and both $t_0$ and $t$ are needed to specify the flow.

An \emph{invariant set} is carried into itself by the dynamics. In a
time-dependent system, the invariant object is usually a moving family
$\mathcal A(t)$ satisfying
\begin{equation}
\varphi_{t,\tau}[\mathcal A(\tau)]=\mathcal A(t).
\label{sm:invariance}
\end{equation}
For example, a trajectory $X_b(t)$ is represented by the family of points
$\mathcal A(t)=\{X_b(t)\}$. A manifold is a curve, surface, or its
higher-dimensional analogue. An \emph{invariant manifold} is a manifold made
of trajectories, so a trajectory starting on it remains on it.

The stable and unstable manifolds of $X_b$ contain the trajectories that
approach it in forward and backward time:
\begin{align}
W_{t_0}^s(X_b)
&=\left\{X_0:\|\varphi_{t,t_0}(X_0)-X_b(t)\|\to0
\text{ as }t\to+\infty\right\},\notag\\
W_{t_0}^u(X_b)
&=\left\{X_0:\|\varphi_{t,t_0}(X_0)-X_b(t)\|\to0
\text{ as }t\to-\infty\right\}.
\label{sm:stable-unstable}
\end{align}
In two-dimensional phase space these manifolds are curves. They can act as
\emph{separatrices}: invariant boundaries between different kinds of motion.

The Letter uses an affine system,
\begin{equation}
\dot X=A(t)X+F(t).
\label{sm:affine}
\end{equation}
The difference $\delta$ between any two solutions satisfies the homogeneous
equation $\dot\delta=A(t)\delta$. Therefore the stable and unstable geometry
of the affine system is determined exactly by its homogeneous part, not by a
small-displacement approximation.

\subsection{Hyperbolic barriers and exponential dichotomy}

For an autonomous one-dimensional barrier, the transition state is a saddle
equilibrium at the barrier top. The inverted harmonic barrier gives the
simplest example:
\begin{equation}
\frac{d}{dt}
\begin{pmatrix}x\\p\end{pmatrix}
=
\begin{pmatrix}0&1/m\\m\omega^2&0\end{pmatrix}
\begin{pmatrix}x\\p\end{pmatrix}.
\label{sm:autonomous-saddle}
\end{equation}
The eigenvalues are $\pm\omega$. The corresponding solutions are
\begin{equation}
X_u(t)=e^{\omega(t-t_0)}
\begin{pmatrix}1\\m\omega\end{pmatrix},
\qquad
X_s(t)=e^{-\omega(t-t_0)}
\begin{pmatrix}1\\-m\omega\end{pmatrix}.
\label{sm:saddle-modes}
\end{equation}
The unstable solution grows forward in time, whereas the stable solution
decays. Their directions form the stable and unstable manifolds of the saddle.

For a time-dependent barrier there is generally no fixed saddle. It is
replaced by a trajectory that remains near the barrier for all time. The
stable and unstable directions cannot usually be found from the eigenvectors
of $A(t)$ at one instant, because those directions need not be preserved as
the system evolves. Instead, stability is defined by the full time evolution
\cite{bartsch_transition_2005,rice_time-dependent_2008}.

Let $\Phi(t,\tau)$ be the propagator of $\dot\delta=A(t)\delta$, so that
$\delta(t)=\Phi(t,\tau)\delta(\tau)$. An \emph{exponential dichotomy} is a
time-dependent splitting into stable and unstable subspaces. A projector
$P(t)$ selects the stable subspace,
\begin{equation}
E^s(t)=\operatorname{Ran}P(t),
\qquad
E^u(t)=\operatorname{Ran}[I-P(t)],
\label{sm:subspaces}
\end{equation}
Every displacement is then the unique sum of one stable component and one
unstable component. The propagator carries these subspaces from one time to
another:
\begin{equation}
\Phi(t,\tau)P(\tau)=P(t)\Phi(t,\tau).
\label{sm:covariant-projector}
\end{equation}
The splitting is exponentially stable if constants $C,\lambda>0$ exist such
that
\begin{align}
\|\Phi(t,\tau)P(\tau)\|
&\le Ce^{-\lambda(t-\tau)}, &&t\ge\tau,\notag\\
\|\Phi(t,\tau)[I-P(\tau)]\|
&\le Ce^{-\lambda(\tau-t)}, &&t\le\tau.
\label{sm:dichotomy}
\end{align}
The first inequality gives forward decay in the stable subspace; the second
gives backward decay in the unstable subspace. This is the time-dependent
counterpart of the stable and unstable eigendirections of a saddle
\cite{coppel_dichotomies_1978,barreira_stability_2008}.

An exponential dichotomy also identifies the trajectory trapped near the
moving barrier \cite{coppel_dichotomies_1978}.

\begin{proposition}[Unique bounded trajectory]
\label{prop:bounded}
If $\dot\delta=A(t)\delta$ has an exponential dichotomy on $\R$ and $F(t)$ is
bounded, then Eq.~\eqref{sm:affine} has one and only one solution bounded for
all $t\in\R$:
\begin{equation}
X_b(t)=\int_{-\infty}^{t}\Phi(t,\tau)P(\tau)F(\tau)\,d\tau
-\int_{t}^{\infty}\Phi(t,\tau)[I-P(\tau)]F(\tau)\,d\tau.
\label{sm:bounded-solution}
\end{equation}
Its stable and unstable manifolds are
\begin{equation}
W_t^{s,u}(X_b)=X_b(t)+E^{s,u}(t),
\qquad
W_t^s(X_b)\cap W_t^u(X_b)=\{X_b(t)\}.
\label{sm:affine-manifolds}
\end{equation}
\end{proposition}

The dichotomy bounds make both integrals converge. Any other bounded solution
would differ from $X_b$ by a homogeneous solution bounded in both time
directions, but an exponential dichotomy allows no such nonzero solution.
Equation~\eqref{sm:affine-manifolds} follows because the stable and unstable
subspaces intersect only at the zero vector. Thus $X_b$ is the time-dependent
replacement for the saddle equilibrium: it is the unique trajectory that
remains at the bottleneck in both time directions.

For the two-dimensional barrier dynamics in the Letter, an invariant line can
also be described by its slope $s=\delta_p/\delta_x$. Substitution into the
homogeneous equations gives
\begin{equation}
\dot s=m\Omega^2(t)-\frac{s^2}{m}.
\label{sm:riccati}
\end{equation}
Solutions of this first-order equation cannot cross while their slopes remain
finite. Therefore an ordering of the stable, unstable, and physical slopes,
once established, is preserved in time.

\subsection{Transition-state geometry}

Three geometric objects have different roles. The \emph{transition state} is
the invariant object trapped at the barrier. Its stable and unstable manifolds
separate nearby reactive and nonreactive trajectories. The \emph{dividing
surface} is the surface crossed when a trajectory moves from reactants to
products. The transition state anchors this surface at the bottleneck
\cite{Pechukas_McLafferty_1973,pollak_transition_1978,Waalkens_Schubert_Wiggins_2008}.

A moving dividing surface can be written as $\sigma(X,t)=0$. A trajectory
crosses it when $\sigma(X(t),t)$ changes sign. For the surface used in the
Letter,
\begin{equation}
\sigma(X,t)=x-x_b(t).
\end{equation}
At a crossing on the physical graph,
\begin{equation}
\frac{d\sigma}{dt}=\frac{p-p_b(t)}{m}=\frac{g(t)}{m}.
\label{sm:crossing-rate}
\end{equation}
If the momentum gap $g(t)$ keeps one sign, every crossing has the same
direction. If each reactive trajectory crosses only once, the surface is
recrossing-free and the flux is exactly one way.

With more than one degree of freedom, motion parallel to the barrier can
continue while the reactive coordinate remains trapped. The transition state
then becomes a \emph{normally hyperbolic invariant manifold} (NHIM). It is
invariant because trajectories starting on it remain on it. It is normally
hyperbolic because perturbations away from it grow or decay faster than motion
within it. Normal hyperbolicity allows the NHIM and its stable and unstable
manifolds to persist under sufficiently small perturbations
\cite{wiggins_normally_1994,wiggins_role_2016}.

In the one-dimensional problem studied in the Letter, there are no internal
directions along the barrier. The NHIM therefore reduces to one point at each
time, or equivalently to the single bounded trajectory $X_b(t)$. The
multidimensional extension would replace this trajectory by a higher-
dimensional trapped set.

\subsection{Lagrangian descriptors}

A Lagrangian descriptor is a numerical tool for locating invariant manifolds.
For each initial condition, it integrates a positive quantity along the
trajectory over a finite time
\cite{Mancho_Wiggins_Curbelo_Mendoza_2013,Lopesino_etal_2017}. The Letter uses
\begin{align}
\mathcal L^+(X_0,t_0)
&=\int_{t_0}^{t_0+T_{\mathrm{LD}}}|p(t;X_0,t_0)|^\gamma\,dt,\notag\\
\mathcal L^-(X_0,t_0)
&=\int_{t_0-T_{\mathrm{LD}}}^{t_0}|p(t;X_0,t_0)|^\gamma\,dt,
\qquad
\mathcal L=\mathcal L^++\mathcal L^- ,
\label{sm:ld}
\end{align}
with $0<\gamma\le1$. Plotting $\mathcal L$ over initial conditions produces
sharp features at separatrices. The forward part $\mathcal L^+$ reveals the
stable manifold, the backward part $\mathcal L^-$ reveals the unstable
manifold, and their intersection locates $X_b(t_0)$
\cite{Revuelta_Benito_Borondo_2021,junginger_transition_2016}.

\subsection{Quantum trajectories and the physical phase-space graph}

Consider a spinless particle satisfying the Schr\"odinger equation
\begin{equation}
i\hbar\,\partial_t\psi
=\left[-\frac{\hbar^2}{2m}\nabla^2+V(x,t)\right]\psi.
\label{sm:schrodinger}
\end{equation}
Where $\psi\ne0$, write the wavefunction in polar form,
\begin{equation}
\psi(x,t)=R(x,t)e^{iS(x,t)/\hbar},
\qquad R\ge0.
\label{sm:polar-form}
\end{equation}
This hydrodynamic representation was introduced by Madelung
\cite{madelung_quantentheorie_1927}.
Substituting this expression into the Schr\"odinger equation and separating
real and imaginary parts gives
\begin{align}
\partial_t\rho+\nabla\!\cdot\!\left(\rho\frac{\nabla S}{m}\right)&=0,
\qquad \rho=R^2,
\label{sm:continuity}\\
\partial_tS+\frac{|\nabla S|^2}{2m}+V+Q&=0,
\label{sm:qhj}
\end{align}
where
\begin{equation}
Q(x,t)=-\frac{\hbar^2}{2m}\frac{\nabla^2R}{R}
\label{sm:qpotential}
\end{equation}
is the quantum potential. These equations are exactly equivalent to the
Schr\"odinger equation wherever the polar form is regular; no semiclassical
approximation has been made
\cite{bohm_suggested_1952,bohm_suggested_1952-1,Holland_1993,
Durr_Teufel_2009,Norsen_2017}.

Equation~\eqref{sm:continuity} is a continuity equation with velocity
\begin{equation}
v(x,t)=\frac{\nabla S(x,t)}{m}.
\label{sm:velocity}
\end{equation}
Its characteristics, or quantum trajectories, satisfy
\begin{equation}
\dot x(t)=v(x(t),t).
\label{sm:guidance}
\end{equation}
An ensemble initially distributed as $|\psi(x,t_0)|^2$ remains distributed as
$|\psi(x,t)|^2$, because both densities obey the same continuity equation.
Thus the trajectories describe the exact transport of Born probability
\cite{Durr_Goldstein_Zanghi_1992}.

Assigning the momentum
\begin{equation}
p(t)=\nabla S(x(t),t)
\end{equation}
lifts each trajectory into phase space. Taking the gradient of
Eq.~\eqref{sm:qhj} along the trajectory gives
\begin{equation}
\dot x=\frac{p}{m},
\qquad
\dot p=-\nabla(V+Q).
\label{sm:newton-form}
\end{equation}
The physical trajectories therefore lie on the time-dependent graph
\begin{equation}
\Gamma_\psi(t)=\{(x,p):p=\nabla S(x,t)\}.
\label{sm:physical-graph}
\end{equation}
This family of graphs is invariant under the phase-space flow
\cite{deLeon_Sardon_2017,RomanRoy_2021}.
This graph assigns one momentum to each position. It represents the quantum
probability current; it does not represent a simultaneous measurement of
position and momentum.

Born probability is supported on $\Gamma_\psi(t)$. The corresponding
phase-space measure is
\begin{equation}
d\mu_t=\rho(x,t)\,\delta\!\left(p-\nabla S(x,t)\right)\,dx\,dp.
\label{sm:physical-measure}
\end{equation}
For any function $h(x,p)$,
\begin{equation}
\int_{\mathcal P}h(x,p)\,d\mu_t
=\int_{\R}h\!\left(x,\nabla S(x,t)\right)\rho(x,t)\,dx.
\label{sm:pushforward}
\end{equation}
The delta function simply restricts the phase-space integral to the physical
graph.

Once $Q(x,t)$ is known, Eq.~\eqref{sm:newton-form} can also be solved from
initial points away from $\Gamma_\psi(t)$. These auxiliary solutions form the
\emph{ambient} phase-space flow. They carry no Born probability for the
prescribed state \cite{wiggins_geometric_2026}, but their stable and unstable
manifolds intersect the physical graph and divide its trajectories into
reactive and nonreactive classes. In particular, the bounded
trajectory $X_b$ belongs to this ambient flow and need not lie on
$\Gamma_\psi(t)$. The momentum gap $g(t)$ measures its separation from the
physical graph.

\section{Gaussian barrier-top model}
\label{sec:gaussian}

\subsection{Gaussian propagation and the effective barrier}

Consider the inverted harmonic potential
\begin{equation}
V(x)=-\frac12m\omega^2x^2
\label{sm:inverted-potential}
\end{equation}
and the Gaussian wavepacket
\begin{equation}
\psi(x,t)\propto
\exp\!\left\{\frac{i}{\hbar}
\left[\mathcal A_t(x-q_t)^2+p_t(x-q_t)\right]\right\}.
\label{sm:gaussian}
\end{equation}
The omitted factor is independent of $x$. The center, momentum, and complex
width evolve as~\cite{Heller_2018}
\begin{align}
q_t&=q_0\cosh(\omega t)
+\frac{p_0}{m\omega}\sinh(\omega t),
\label{sm:qdef}\\
p_t&=p_0\cosh(\omega t)+m\omega q_0\sinh(\omega t),
\label{sm:pdef}\\
\mathcal A_t
&=\frac{m\omega}{2}
\frac{2\mathcal A_0\cosh(\omega t)+m\omega\sinh(\omega t)}
     {2\mathcal A_0\sinh(\omega t)+m\omega\cosh(\omega t)}.
\label{sm:Adef}
\end{align}
Write $\nu(t)=\operatorname{Im}\mathcal A_t$. For a normalizable packet,
$\nu(t)>0$ at every finite time. The amplitude is
$R\propto\exp[-\nu(t)(x-q_t)^2/\hbar]$, which gives
\begin{equation}
Q(x,t)=\frac{\hbar\nu(t)}{m}
-\frac{2\nu^2(t)}{m}(x-q_t)^2.
\label{sm:gaussian-Q}
\end{equation}
Define
\begin{equation}
a(t)=\frac{4}{m}\nu^2(t).
\label{sm:adef}
\end{equation}
Apart from a term independent of $x$, the effective potential is
\begin{equation}
U_{\mathrm{eff}}(x,t)=V+Q
=-\frac12m\omega^2x^2-\frac{a(t)}{2}(x-q_t)^2.
\label{sm:ueff}
\end{equation}
Its curvature is
\begin{equation}
\partial_x^2U_{\mathrm{eff}}
=-\left[m\omega^2+a(t)\right]
\equiv-m\Omega^2(t),
\qquad
\Omega^2(t)=\omega^2+\frac{a(t)}{m}.
\label{sm:curvature}
\end{equation}
Because $a(t)\ge0$, the quantum potential preserves the barrier:
$\Omega^2(t)\ge\omega^2>0$.

\subsection{Barrier top and forcing}

The maximum of $U_{\mathrm{eff}}$ is obtained from
$\partial_xU_{\mathrm{eff}}=0$:
\begin{equation}
x_*(t)=\frac{a(t)q_t}{m\Omega^2(t)}.
\label{sm:xstar}
\end{equation}
The quantum potential therefore shifts the effective barrier top toward the
wavepacket center. In the original coordinate $x$, the force is
\begin{equation}
-\partial_xU_{\mathrm{eff}}
=m\Omega^2(t)x+f(t),
\qquad
f(t)=-a(t)q_t.
\label{sm:forcing}
\end{equation}
This gives the affine phase-space equation used in the Letter,
\begin{equation}
\frac{d}{dt}\begin{pmatrix}x\\p\end{pmatrix}
=
\begin{pmatrix}0&1/m\\m\Omega^2(t)&0\end{pmatrix}
\begin{pmatrix}x\\p\end{pmatrix}
+\begin{pmatrix}0\\f(t)\end{pmatrix}.
\label{sm:model-affine}
\end{equation}
Because $U_{\mathrm{eff}}$ is exactly quadratic, this affine equation is exact
for the Gaussian model.

\subsection{Decay of the Gaussian forcing}
\label{sm:decay}

The Letter requires $f(t)$ to be bounded and integrable. These properties
follow directly from the Gaussian width.

\begin{proposition}
Let $\mathcal A_0=\mu_0+i\nu_0$ with $\nu_0>0$. Then
\begin{equation}
\nu(t)=O\!\left(\operatorname{sech}^2\omega t\right),
\qquad
a(t)=O\!\left(\operatorname{sech}^4\omega t\right),
\qquad
f(t)=O\!\left(\operatorname{sech}^3\omega t\right).
\label{sm:decay-bounds}
\end{equation}
Consequently, $a(t)\to0$ and $f(t)\to0$ as $|t|\to\infty$, and
$f\in L^1(\R)\cap L^\infty(\R)$.
\end{proposition}

\begin{proof}
Taking the imaginary part of Eq.~\eqref{sm:Adef} gives
\begin{equation}
\nu(t)=
\frac{\nu_0m^2\omega^2\operatorname{sech}^2(\omega t)}
{G(\tanh\omega t)},
\label{sm:nu-decay}
\end{equation}
where
\begin{equation}
G(T)=4|\mathcal A_0|^2T^2+4\mu_0m\omega T+m^2\omega^2.
\end{equation}
Completing the square shows that
\begin{equation}
G(T)\ge
\frac{m^2\omega^2\nu_0^2}{|\mathcal A_0|^2}>0.
\end{equation}
Therefore $\nu(t)=O(\operatorname{sech}^2\omega t)$. Since
$a=4\nu^2/m$, the second estimate follows. Equations~\eqref{sm:qdef} and
$|\sinh y|\le\cosh y$ give $|q_t|=O(\cosh\omega t)$. Hence
$|f|=|a q_t|=O(\operatorname{sech}^3\omega t)$. The function
$\operatorname{sech}^3(\omega t)$ is bounded and integrable on $\R$, which
proves the result.
\end{proof}

For the state used in the Letter,
$\mathcal A_0=\tfrac12m\omega(1+i)$, so $\nu_0>0$. The estimates above verify
the decay and boundedness assumptions used for the affine forcing. They also
show that $\Omega^2(t)\to\omega^2$ as $|t|\to\infty$, while
$\Omega^2(t)\ge\omega^2>0$ at every time. Together with the dichotomy argument
in the Letter, these facts yield the unique bounded transition-state
trajectory of Proposition~\ref{prop:bounded}.

\section{Accompanying source code}

The plain-text file
\texttt{quantum\_transition\_state\_figures.txt} accompanying this
Supplemental Material contains the Python source code used to generate the
transition-state structure and Lagrangian-descriptor figures in the Letter.
Installation requirements and execution instructions are provided in
\texttt{README.TXT}.

\bibliography{quantum_transition_state}